\documentclass{ws-mpla}

\usepackage{amsmath}

\begin{document}

\markboth{J. Santiago}
{Bounds on the String Scale from Flavor Physics}

%
\catchline{}{}{}{}{}
%

\title{
NEW BOUNDS ON THE STRING SCALE \\
FROM FLAVOR PHYSICS
}

\author{\footnotesize J. SANTIAGO
}

\address{IPPP, University of Durham,\\
Durham, DH1 3LE, UK
\\
jose.santiago-perez@durham.ac.uk}

\maketitle

\pub{Received (Day Month Year)}{Revised (Day Month Year)}

\begin{abstract}
We review the very stringent lower bounds on the string scale that
arise from flavor considerations in models with intersecting
branes. Despite the absence of a realistic flavor structure
at tree level, flavor changing interactions induce a non-trivial
pattern of fermion masses and mixing  angles when quantum corrections
are taken into account. The resulting realistic theory of flavor
allows us to constrain, in an unambiguous way, the string scale up to
levels difficult to reconcile non-supersymmetric models.

\end{abstract}

\ccode{PACS Nos.: include PACS Nos.}

\section{Introduction}

After the so-called second string revolution, string phenomenology has
widen its subject beyond heterotic models. One very exciting
possibility came across a few years ago through the realization that
D-branes intersecting at non-trivial angles~\cite{Berkooz:1996km} 
allow for a neat way of
breaking supersymmetry~\cite{Bachas:1995ik},
generating four-dimensional chirality, and
constructing in a bottom-up approach string models with a fully
realistic matter and symmetry content~\cite{Ibanez:2001nd}. 
A lot of effort has been put
since in the study of the model building and phenomenology of such
theories. (See~\cite{Blumenhagen:2000wh} for the initial works
and~\cite{Abel:stringreview} for a more detailed introduction and
references.) This intense dedication has resulted almost futile regarding
the search of supersymmetric models. The restrictions from
Ramond-Ramond tadpole cancellation are so strict that when added to
the requirement of the preservation of $N=1$ supersymmetry the
resulting models are scarce and usually not fully
realistic~\cite{Cvetic:2001tj}, with extra exotic
matter in the spectrum. (See however important progress in
this direction in Ref.~\cite{Honecker:2003vq}.) 

Given the fact that most of the realistic models presented so far are
non-supersymmetric, a sensible question to ask is whether low scale
non-fine-tuned models are phenomenologically allowed. A necessary
previous step, to which we devote the first half for the present review, 
is the investigation of the structure of flavor in these models. The
trivial structure of tree level Yukawa couplings present in realistic 
models~\cite{Cremades:2003qj}
might seem discouraging at first sight. 
However the presence of flavor 
violating processes~\cite{Abel:2003fk}, 
both at the string and field theory levels,
propagates through quantum corrections to give a rich, realistic
structure of fermion masses and mixing angles when one loop
contributions are taken into account.  

Armed with this realistic flavor structure we can study, in a
quantitative unambiguous way, the phenomenological implications of low
scale intersecting brane models. The same flavor violating four-point
amplitudes that give raise to a realistic fermion spectrum induce
important tree-level contributions to meson oscillations and rare
processes. These are extremely well constrained experimentally,
implying very stringent bounds on the string scale. (Incidentally, we
should mention here that while tree level contributions to flavor
violating processes decouple as the string scale increases, their
contribution to the one loop Yukawa couplings does not.) In an attempt
to be as comprehensive as possible and in order to minimize the
possibility of fine-tuned situations in which the flavor violating
contributions are small, we also discuss a new host of bounds on these
models arising from flavor conserving amplitudes. Although the
resulting constraints are much milder, still a lower bound on the
string scale in the tens of TeV region is obtained.

\section{Flavor Structure of Models with Intersecting Branes}

Models with intersecting branes have a number of very interesting
features such as the presence of chiral fermions, family replication and
the possibility of constructing models with precisely the symmetries
and the matter content of the Standard Model. Once a model has
successfully passed such a coarse graining sift one has to
more finely test it by for instance checking whether a realistic
pattern of fermion masses and mixing angles can arise and what
phenomenological predictions and restrictions it has.
Two main ingredients enter the flavor
structure of these models, one is the form of the Yukawa couplings,
the other the presence of tree level flavor changing four point
amplitudes. These two features are intimately related for, as we shall
see, flavor violating operators are needed to generate a fully
realistic pattern of fermion masses and mixings but at the same time,
bounds on the string scale from the contribution of these operators to
rare processes cannot be computed unless the flavor structure is
known.

\begin{figure}[th]
\centerline{\psfig{file=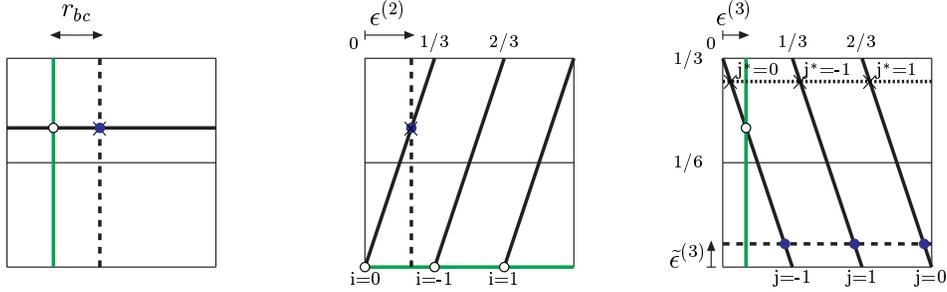,width=12.5cm}}
\vspace*{8pt}
\caption{Brane configuration in a model of D6-branes intersecting at angles.
The leptonic sector is not represented while the baryonic, left, right and
orientifold image of the right are respectively the dark solid, faint solid,
dashed and dotted. The intersections corresponding to the quark
doublets ($i=-1,0,1$), 
up type singlets ($j=-1,0,1$) and down type singlets ($j^\ast=-1,0,1$)
are denoted by 
an empty circle, full circle and a cross, respectively.
All distance parameters are measured in units
of $2 \pi R$ with $R$ the corresponding radius (except
$\tilde{\epsilon}^{(3)}$ which is measured 
in units of $6 \pi R$)\label{figuratoros}}
\end{figure}

In order to make our arguments more specific, we shall concentrate on
a particular model represented in 
Fig.~\ref{figuratoros}~\cite{Cremades:2003qj}. 
It corresponds to an orientifold compactification of type II A
theory with four stacks of D6-branes wrapping factorizable 3-cycles on
the compact dimensions. The compactified space is a 
factorizable 6-Torus 
$T^2\times T^2 \times T^2$,
and the orientifold projection is given by $\Omega \mathcal{R}$ where
$\Omega$ is the world-sheet parity and $\mathcal{R}$ is a reflection
about the horizontal axis of each of the three 2-tori,
\[
\mathcal{R} Z_I = \bar{Z}_I.
\]
We have denoted the coordinates of the tori by complex coordinates
$Z_I=X_{2I+2}+\mathrm{i} X_{2I+3}$, $I=1,2,3$, 
so the three boxes in the figure represent each 2 torus, 
with opposite edges being identified. 
The branes have four extended dimensions plus three
compactified ones. Therefore, they 
appear as just lines in each $T_2$. The net 
effect of the orientifold projection is to introduce mirror images of
the branes in each $T_2$ (in the plane running horizontally). 

The matter content of the low energy spectrum 
consists of massless chiral 4-dimensional fermions plus generally
massive scalars, both living at branes intersections and transforming
as bi-fundamentals of the corresponding gauge groups. The masses of
the latter depend on the particular brane configuration (angle between
branes) and can be considered as the superpartners of the fermions
(for specific values of the angles in which they become massless, some
$N=1$ supersymmetry will be preserved by the corresponding
intersection). Gauge bosons live in the world-volume of the branes,
corresponding to unitary (orthogonal or simplectic groups are also
possible for orientifold compactifications) gauge groups.
Our particular model
contains at low energies just the particle content and
symmetries of the MSSM. In order to get that, the model contains 
four stacks of
D6-branes, called \textit{baryonic} (a), \textit{left} (b), \textit{right}
(c), and \textit{leptonic} (d). Three of the dimensions of each D6-brane
wrap a 1-cycle on each of the three 2-tori, with wrapping numbers
denoted by $(n^I_k,m^I_k)$, \textit{i.e.} the stack $k$ wraps $n^I_k$ times
the horizontal dimension of the $I-$th torus and $m^I_k$ times the
vertical direction. We have to include for consistency their orientifold
images with $(n^I_k,-m^I_K)$ wrapping numbers. The number of
branes in each stack, their wrapping numbers and the gauge groups
they give rise to are shown in Table~\ref{data:branes:model} and a
subset of them, together with some of the relevant moduli, are displayed
in Fig.~\ref{figuratoros}.
\begin{table}[h]
\tbl{
Number of branes, gauge groups and wrapping numbers for
the different stacks in the model discussed in the text.
\label{data:branes:model}}
{
\begin{tabular}{@{}clccc@{}}
\toprule
Stack & Name &  $N_k$ & Gauge group & wrapping numbers
\\
\colrule
a & \textit{baryonic} & 3 & $\mathrm{SU}(3)\times \mathrm{U}(1)_a$
&(1,0);(1,3);(1,-3) 
\\
b & \textit{left} & 1 & $\mathrm{SU}(2)$ & (0,1);(1,0);(0,-1)
\\
c & \textit{right} & 1 & $ \mathrm{U}(1)_c$ & (0,1);(0,-1);(1,0)
\\
d & \textit{leptonic} & 1 & $\mathrm{U}(1)_d$ &(1,0);(1,3);(1,-3)
\\
\botrule
\end{tabular}
}
\end{table}

A crucial feature that governs the whole flavor structure of these
models is the fact that different families as well as the Higgs boson
live at separate points in the compact dimensions. Yukawa interactions
then correspond to instantonic contributions that are therefore
expected to be proportional to the exponential of the relevant area
connecting the three vertices. A more detailed study of Yukawa couplings,
using calibrated geometry~\cite{Cremades:2003qj}, and confirmed later
by a proper string calculation using 
conformal field theory techniques~\cite{Cvetic:2003ch,Abel:2003vv}, 
showed that when the compact space
is a factorizable torus and the branes wrap factorizable cycles,
the relevant area is the sum of the \textit{projected} areas of the triangle
over each sub-torus. The final result, including the quantum part reads
\begin{equation}
Y=\sqrt{2} \lambda_{II} 2 \pi \sum_{I=1}^3 \sqrt{\frac{4\pi B(\nu_I,1-\nu_I)}
{B(\nu_I,\theta_I)F(\nu_I,1-\nu_I-\theta_I)}}\sum_m
\mathrm{e}^{-\frac{A_I(m)}{2\pi \alpha^\prime}},
\end{equation}
where we have neglected the presence of non-zero $B$ field and
Wilson lines and $\lambda_{II}$ is the string coupling, 
$B$ is the Euler Beta function, $I$ runs over the three tori,
$\nu_I$ and $\theta_I$ are the angles at the fermionic intersections,
$m$ runs over all possible triangles connecting the three vertices
on each of the three tori (there is an infinite number of them due
to the toroidal periodicity) and $A_I(m)$ is the projected area of the
$m-$th triangle on the $I-$th torus.

Another important feature of the model is the fact that family
\textit{separation} occurs at a different torus for each chirality. As
can be seen in Fig.~\ref{figuratoros}, left handed fermions are split
apart in the second torus (called hereafter \textit{left} torus)
whereas they are localized at the same point in the third torus (the
\textit{right} one). The opposite happens for the right handed
fields. This fact, together with the previously mentioned that the
couplings are proportional to the exponential of minus the
\textit{projected} areas, result in a trivial, factorizable form of
the Yukawa couplings at tree level
\begin{equation}
(Y_u^{tree})_{ij}\sim a_i b^u_j,
\quad
(Y_d^{tree})_{ij}\sim a_i b^d_j.
\end{equation}
The resulting spectrum is clearly unrealistic, with one massive and
two massless generations. Nonetheless we will shortly see that
the same feature of fermion splitting in the compact dimensions
originates flavor violating four-point amplitudes that will contribute
at loop level to give a non-trivial, fully realistic structure to the
Yukawa couplings.

It has been known for long that models with split fermions suffer from
flavor changing neutral current problems arising from the
family non-universal couplings of the gauge boson KK
modes~\cite{Delgado:1999sv}. A 
detailed study of the four-point amplitudes~\cite{Abel:2003vv}  
in these models uncovered new sources of flavor violation mediated by
string instantons~\cite{Abel:2003fk,Abel:Lebedev:Santiago}. These can
be chirality preserving, whose contribution is again proportional to
the exponential of the area of the corresponding
quadrangle (with a non-zero area in only one torus for the leading
contribution) or chirality changing, with a much richer structure and
relevance regarding the generation of non-trivial one loop Yukawa
couplings. It is the latter that we turn our attention to now.

\begin{figure}[th]
\centerline{\psfig{file=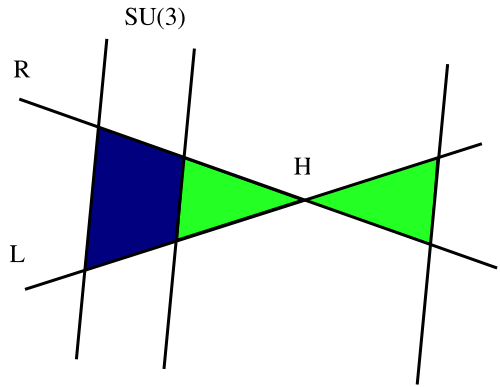,width=6.5cm}}
\vspace*{8pt}
\caption{
Higgs-like processes mediated by the Higgs boson (green) and string
instantons (blue). The former is proportional to the product of Yukawa
couplings whereas the latter introduce new flavor structure.
\label{crossvsnoncross}}
\end{figure}

We are interested in the following chirality changing four-point
amplitudes (for concreteness we consider here quark fields as external
particles)
\[
\mathcal{A}^{LR}_{ijkl} (\bar{q}_{L_i} q_{R_j}) (\bar{q}_{R_k} q_{L_l}).
\]
At the field theory level, such amplitude can be originated by the
exchange of the Higgs boson. This process is indeed reproduced in the
string calculation, together with other processes, purely stringy in origin,
that will be crucial to generate a non-trivial fermionic spectrum.
For the sake of the discussion we concentrate for the moment on one
particular sub-torus. The relevant amplitude corresponds to a
four-sided polygon with two left handed and two right handed quarks as
vertices. There are two possibilities for such a polygon, displayed in
Fig.~\ref{crossvsnoncross}, one is obtained by adding together two
triangular areas joined by the Higgs vertex (green area in the
figure), the string calculation agrees with the intuition, giving a
contribution proportional to the product of the two Yukawa couplings
(one per triangle) and with a Higgs pole structure,
\begin{equation}
\mathcal{A}^{LR}_{ijkl} \sim \frac{Y_{ij}Y_{kl}^\dagger}{t-M_H^2},
\end{equation}
where we have shown the $t$-channel exchange for the sake of the
example.
The second possibility, represented in blue in the figure, 
lacks a field theory counterpart. In the case that there is a non-zero
area in just one torus, this contribution is again proportional to the
exponential of the area swept out by the string connecting the four
vertices, this time without running through a Higgs vertex and
therefore being suppressed by the string scale instead of having a
Higgs pole like the previous contribution. Two important effects are
originated from this kind of diagrams, the first is already apparent
now. Even in the case of non-zero area in just one torus, although the
amplitude is factorizable, it is not proportional to the Yukawa
couplings, therefore introducing new, yet too trivial to
generate a fully realistic spectrum, flavor structure in the game. The
second feature, this time enough to realize a realistic fermion spectrum
happens when there are non-zero areas in more than one torus. In that
case, as we shall see in a moment, factorization of the amplitude is
lost and fully non-trivial new flavor structure is generated.

The technical reason for this non-factorization is the following. In
order to perform the string calculation one maps the world-sheet into
the upper complex plane, with the vertices in the real axis. Three of
the vertices are conventionally fixed at positions 0, 1 and  $\infty$
using $SL(2,R)$ invariance 
whereas one has to integrate over the position of the last one, that
is situated at $0<x<1$. The classical contribution is then obtained by
minimizing the action with respect to this parameter $x$ and
performing a saddle point approximation. The crucial point is that
there is only \textit{one} parameter $x$ for \textit{all} the three
tori and although the minimization of the action for each torus gives
the projected area in that torus, the value of $x_{min}$ is usually
not the same on the three of them and therefore the minimum of the total
action does not correspond to the separate minimum of each sub-factor.
Apart from the trivial cases in which the areas are equal up to
rescaling in all tori or there is a non-zero area in just one torus, 
there are two (related) cases in which there
is still factorization. One corresponds to three point amplitudes that
give rise to Yukawa couplings, they are proportional to the projected
areas and therefore factorize because we can fix the three vertices
and there is not any extra free parameter that could make a difference
between the different tori. The second case is chirality flipping
amplitudes mediated by the Higgs field. In that case, the minimum of
the action corresponds to $x_{min}=1$, which is of course common to
all tori.

\begin{figure}[ht]
\centerline{\psfig{file=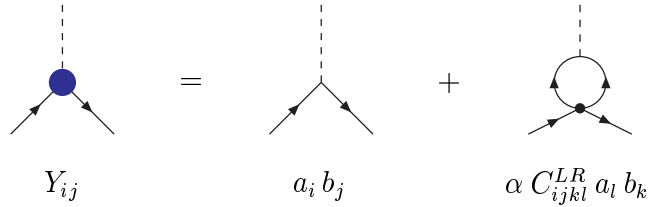,width=8.5cm}}
\vspace*{8pt}
\caption{
Structure of Yukawa couplings including one loop threshold
effects. The black dot corresponds to tree level chirality changing
couplings. 
\label{oneloopyuk}}
\end{figure}

We have seen that the tree level Yukawa couplings are factorizable and
therefore too trivial for a realistic spectrum. Chirality and flavor
changing amplitudes have been found of three different classes flavor
wise, one is
mediated by the Higgs boson, they are not only factorizable but also
proportional to the tree level Yukawa couplings. The second class is
that of factorizable but not proportional to Yukawa couplings
contributions, these are for instance the ones in which there is a
non-zero area in just one sub-torus. Finally there are other
contributions that do not factories, corresponding for instance 
to non-zero area of
the blue type in Fig.~\ref{crossvsnoncross} in more than one torus.
The very rich structure present in four-point amplitudes propagates
through quantum corrections to the Yukawa couplings. Even though a
full string calculation is possible, we will treat threshold
corrections at the field theory level, plugging the values of the
chirality changing amplitudes we have just computed as effective
vertices. The one loop corrected Yukawa couplings have then a flavor
dependence shown in Fig.~\ref{oneloopyuk}. Corrections proportional to
the Yukawa couplings give no new structure but just renormalizes the
tree level values, factorizable but non proportional to Yukawa
corrections increase the rank of the Yukawa matrix in one unit,
therefore giving mass to the second generation. Finally,
non-factorizable corrections give masses to all the three generations
allowing therefore for a fully realistic spectrum. In fact, the
different contributions we have just mentioned, give a rationale for
the hierarchical spectrum observed in nature. Masses for the third
family are generated at tree level, thus it is natural for them to be
large. The first two generations, getting masses at one loop, are
naturally much lighter than the third one. Furthermore, an extra
hierarchy can naturally arise if the non-factorizable corrections are
suppressed with respect to the factorizable ones.  In order to test in
a quantitative way our assertions, we have performed a fit to the
quark masses and mixing angles reproducing reasonably well the
observed spectrum and obtaining in this way a well defined flavor
pattern that will allow us to give unambiguous predictions for the
phenomenological implications of models with intersecting branes.

The relevant (for flavor) parameters defining our model are, the
horizontal and vertical ratios in the second and third torus, the
position of the right brane in the second, $\epsilon^{(2)}$ and the
third torus, $\tilde{\epsilon}^{(3)}$, the position of the left brane
in the third torus, $\epsilon^{(3)}$, and the ratio of the up-type to
down-type Higgs vevs~\footnote{As we mentioned the low energy matter
  content is that of the MSSM, including two Higgs doublets.}, 
$\tan \beta$. A reasonable value of
the quark masses and mixing angles,
\begin{align}
&m_u\sim 4\times 10^{-3} \mbox{ GeV},\quad  m_c\sim 1.8\mbox{ GeV},
\quad \; m_t\sim 176 \mbox{ GeV}, \nonumber \\
&m_d\sim 4\times 10^{-3} \mbox{ GeV},\quad m_s\sim 0.04\mbox{ GeV},
\quad  m_b\sim 8 \mbox{ GeV},\label{massesandmixings:ourresult}\\
& V_{12}\sim 0.22,\,\quad \quad \quad\quad \quad V_{13}\sim 0.003,\quad
\quad \quad V_{23}\sim 0.02,
\nonumber
\end{align}
where we have included a global normalization factor 0.95 in the Yukawa
couplings, 
is obtained for the following parameters (dimensional parameters are
expressed in string units)
\begin{align}
&R^{(2)}_1=1.1, \quad
R^{(3)}_1=1.15, \quad
\chi_2=1.24,\quad
\chi_3=0.94,\quad
\nonumber \\
&\epsilon_2=0.121,\quad
\epsilon_3=0.211,\quad
\tilde{\epsilon}_3=0.068,\quad
\tan\beta=20. \label{our:parameters}
\end{align}
In this particular example, the full flavor structure is fixed, all
the values of the relevant moduli, Yukawa couplings, rotation
matrices, etc. are known in terms of the string scale. 
We can therefore compute without any ambiguity the values of any
observable as a function of the string scale and obtain in that way
precise bounds on it.

We would like to make one further remark before discussing the bounds
on the string scale in these models. As we have said, the vast
majority of the realistic models with intersecting branes presented in
the literature so far are non-supersymmetric. It is to those that the
specific mechanism for non-trivial fermion mass generation through
Yukawa threshold effects apply. One should however consider the
possibility of supersymmetric realistic models (as we will see in the
next section the very constraining bounds on the string scale in these
models makes it highly desirable for them to be supersymmetric) in
which case non-renormalization theorems prevent threshold effects in
Yukawa couplings. The very model we have presented possess a
supersymmetric spectrum if the ratio of horizontal to vertical radii
is equal in the second and third tori (hidden branes that break all
the supersymmetries are anyway present in this model so even though the
contribution we have computed would vanish in the ``supersymmetric
configuration'', there would be similar ones from strings stretching
between the hidden and the MSSM sectors). Nevertheless, even in fully
supersymmetric configurations, there are still sources of non-trivial
flavor structure that can potentially lead to a realistic fermion
spectrum. One possible source is for instance soft breaking A-terms. 
They have a different flavor structure than Yukawa couplings and this
could be propagated through supersymmetric quantum loops to the Yukawa
couplings. A detailed study of this and other effects is still
necessary  to see if realistic flavor patterns can be obtained as well
in supersymmetric models.

\section{Bounds on the String Scale}

Once a realistic pattern of fermion masses and mixing angles has been
developed we can compute, in a well-defined way, the effects of
four-point string amplitudes in different observables in order to
establish lower bounds on the string scale. The presence of flavor
changing processes makes it more likely that rare processes are the
most constraining ones. Nevertheless we will also consider the
contribution to flavor conserving processes that give independent yet
quite strict bounds on the string scale. We start our discussion with
the former.

\subsection{Flavor Violating Observables}

Tree level flavor violating four-point amplitudes directly contribute
to rare processes like meson oscillations and rare lepton and meson
decays. These processes are loop and sometimes GIM suppressed in the
Standard Model and therefore very small. They are also extremely well
constrained experimentally what makes of them a very powerful probe of
flavor violating physics beyond the Standard Model. For this analysis,
we have closely followed the discussion in Ref.~\cite{Langacker:2000ju},
where the phenomenological implications of an extra $U(1)$ gauge boson
with family non-universal couplings was considered. Of course, it has
been extended to take
into account not only the effects of gauge boson KK modes but also
those of string instantons. Instead of giving a full account of the
effect in each of the observables that have been considered, we will discuss
in some detail the case of meson oscillations and refer
to~\cite{Langacker:2000ju,Abel:Lebedev:Santiago} for a more in-depth  
discussion. 

The mass splitting for a meson with quark content $P_0=\bar{q}_j q_i$,
in the vacuum insertion approximation, reads,
\begin{equation}
\Delta m_P = \frac{2m_P F_P^2}{M_s^2} 
\left\{ 
\frac{1}{3} \mathrm{Re} \left[ A^{LL}_{ijij}+ A^{RR}_{ijij} \right]
- 
\left[\frac{1}{2}+\frac{1}{3} \left( \frac{m_P}{m_{q_i}+m_{q_j}}
  \right)^2 \right] \mathrm{Re} A^{LR}_{ijij} \right\},
\end{equation}
where $m_P$ and $F_P$ are, respectively the mass and decay constant of
the meson. Here $A_{ijkl}$ are the dimensionless coefficients
parameterizing the four fermion operators (with $1/M_s^2$ factored
out).

Indirect CP violation in the Kaon system, which has been measured as
well with extreme accuracy, is parametrized by
\begin{equation}
|\epsilon_K| = \frac{m_K F_K^2}{\sqrt{2} \Delta m_K M_s^2} 
\left\{ 
\frac{1}{3} \mathrm{Im} \left[ A^{LL}_{dsds}+ A^{RR}_{dsds} \right]
- 
\left[\frac{1}{2}+\frac{1}{3} \left( \frac{m_K}{m_{d}+m_{s}}
  \right)^2 \right] \mathrm{Im} A^{LR}_{dsds} \right\}.
\end{equation}

We compute the limits on the string scale requiring that the new
contribution is smaller that the experimental value. This implies the
following restrictions on the different coefficients,
\begin{itemize}
\item Kaon mass splitting
\begin{equation}
\frac{1}{M_s^2} | \mathrm{Re} \left[ A^{LL}_{dsds}+ A^{RR}_{dsds}
\right] - 17.1 \mathrm{Re} \left [ A^{LR}_{dsds} \right] | 
\lesssim 3.3 \times 10^{-7} \mbox{ TeV}^{-2},
\end{equation}

\item $B$ mass splitting
\begin{equation}
\frac{1}{M_S^2} 
\left|
\mathrm{Re}\big[ A^{LL}_{dbdb} + A^{RR}_{dbdb}
  \big]
-3\, \mathrm{Re}\big[ A^{LR}_{dbdb} \big]
\right|
\lesssim 2 \times 10^{-6} \mbox{ TeV}^{-2},
\end{equation}
\item $B_s$ mass splitting
\begin{equation}
\frac{1}{M_S^2} 
\left|
\mathrm{Re}\big[ A^{LL}_{sbsb} + A^{RR}_{sbsb}
  \big]
-3\, \mathrm{Re}\big[ A^{LR}_{sbsb} \big]
\right|
\lesssim 6.6 \times 10^{-5} \mbox{ TeV}^{-2},
\end{equation}
\item $D$ mass splitting
\begin{equation}
\frac{1}{M_S^2} 
\left|
\mathrm{Re}\big[ A^{LL}_{ucuc} + A^{RR}_{ucuc}
  \big]
-3.9\, \mathrm{Re}\big[ A^{LR}_{ucuc} \big]
\right|
\lesssim 3.3 \times 10^{-6} \mbox{ TeV}^{-2},
\end{equation}
\item Kaon CP violation
\begin{equation}
\frac{1}{M_S^2} 
\left|
\mathrm{Im}\big[ A^{LL}_{dsds} + A^{RR}_{dsds}
  \big]
-17.1\, \mathrm{Im}\big[ A^{LR}_{dsds} \big]
\right|
\lesssim 2.6 \times 10^{-9} \mbox{ TeV}^{-2}.
\end{equation}
\end{itemize}
It is clear  that if the coefficients of the 4 fermion operators
are order one (times $1/M_S^2$), the Kaon system  puts a constraint 
$M_S \gtrsim 10^{3-4}$ TeV. Using the specific values of the rotation
matrices found before we show in Table~\ref{boundsFCNC:table} that
these estimations are indeed right.

The list observables is completed with $\mu-e$ coherent conversion in
atoms, tau decays, and leptonic and semi-leptonic meson decays. Having
developed a theory of fermion masses and mixing angles for the
quark sector, we can only make an estimation of the bounds implied by
these leptonic observables. Nevertheless, the
order of magnitude of the bounds on the string scale implied by those
is similar of the one found for quark observables, as shown in
Table~\ref{boundsFCNC:table}.

\begin{table}[h]
\tbl{
Bounds on the string scale from flavor violating
  observables.
\label{boundsFCNC:table}
}
{\begin{tabular}{@{}cccc@{}} \toprule
Quark Observables & $M_s \mbox{ (TeV)}
\lesssim$ & (Semi)leptonic Observables &
 $M_s \mbox{ (TeV)} \lesssim$ \\
\colrule
$\Delta m_K$ & $1400$ & $\mu-e$ conversion & 1000 \\
$\Delta m_B$ & $800$ & $\tau\to 3 e$ & 2 \\
$\Delta m_{B_s}$ & $450$ & $\tau\to 3\mu$  & 2 \\
$\Delta m_D$ & $1100$ & $K^0_L\to\mu^+\mu^-$  & 260 \\
$|\epsilon_K|$ & $4\times 10^4$ & $K^0_L\to \pi^0 \mu^+ \mu^-$ & 300
\\ 
\botrule
\end{tabular}}
\end{table}

\subsection{Flavor Preserving Observables} 

Flavor violating observables have proved extremely constraining in
models with intersecting branes. The bounds obtained in the previous
section should be considered as typical bounds for these models. The
long list of observables we have studied makes it highly implausible that a
choice of parameters (moduli) can be made for which \textit{all} the
flavor violating observables are strongly suppressed. In any case, in
order to try and avoid that situation and for the sake of completeness
we consider now the effects on flavor preserving observables. These
effects rely more on very generic features of models with intersecting
branes than in the particular structure of flavor. These features are,
among others, the presence of several $U(1)$ gauge fields and the
appearance of right handed neutrinos. For that reason we shall not use
the detailed form of the rotation matrices to perform a very precise 
computation but will just estimate the effects and the corresponding
bounds on the string scale.

We will start with a flavor conserving but CP violating observable,
electric dipole moment (EDM) of (mainly mercury) atoms. EDM
measurements mostly constrain operators of the type 
$\bar{d}_L d_R \bar{s}_R s_L$, $\bar{d}_L d_R \bar{b}_R b_L$, etc. Let
us concentrate on the first transition for definiteness. The relevant
Lagrangian is therefore
\begin{eqnarray} 
\mathcal{L}&=&\mathcal{A}^{LR}_{ddss} \bar{d}_L d_R \bar{s}_R s_L +
\mathrm{h.c.} \nonumber \\
&=& -\frac{1}{2}\mathrm{Im}\mathcal{A}^{LR}_{ddss} 
(\bar{d} \mathrm{i} \gamma_5 d \bar{s} s
- \bar{s} \mathrm{i} \gamma_5 s \bar{d}d) + \mbox{ CP conserving
  part}.
\end{eqnarray}
Using the recent analysis in Ref.~\cite{Demir:2003js} we can put the
following bound on the corresponding coefficient,
\begin{equation}
\mathrm{Im} \mathcal{A}^{LR}_{ddss} < 6 \times 10^{-11} \,
\mathrm{GeV}^{-2},
\end{equation}
from the mercury EDM. The bounds on the other coefficients and the one
coming from the neutron EDM are weaker than this one.
Including typical values for the mixing angles we obtain the following
(conservative) estimation for the bound on the string scale
\begin{equation}
M_s \gtrsim 10 \, \mathrm{TeV}.
\end{equation}

The next effect we are going to consider is supernova cooling. For
that we will use two rather generic features of models with
intersecting branes, the presence of right handed neutrinos and the
appearance of new $U(1)$ gauge groups that get masses (and therefore
mix among themselves) through a generalized Green-Schwarz
mechanism. The emission of right-handed neutrinos during supernova
collapse can affect the rate of cooling. This implies the following
constraint, for one species of light Dirac neutrinos,
\begin{equation}
\Lambda \gtrsim 200 \, \mathrm{TeV},
\end{equation}
where $\Lambda$ is the scale of the following quark-neutrino interaction
\begin{equation}
\mathcal{L}_R = \frac{4\pi}{\Lambda^2} \bar{q} \gamma_\mu \gamma_5 q
\, \bar{\nu}_R \gamma^\mu \nu_R.
\end{equation}

\begin{figure}[h]
\caption{\label{supernova}
Tree (left) and one loop (right) level emission of right--handed
neutrinos relevant for supernova cooling.}
\begin{center}\includegraphics[%
  scale=0.8]{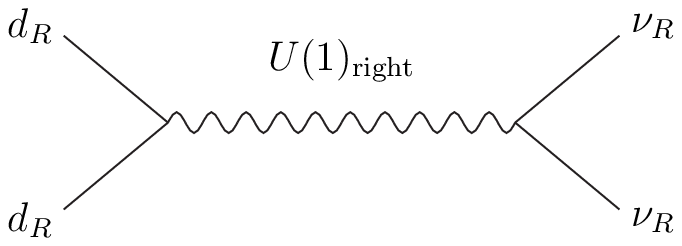}
\hspace{.5cm}
\includegraphics[%
  scale=0.6]{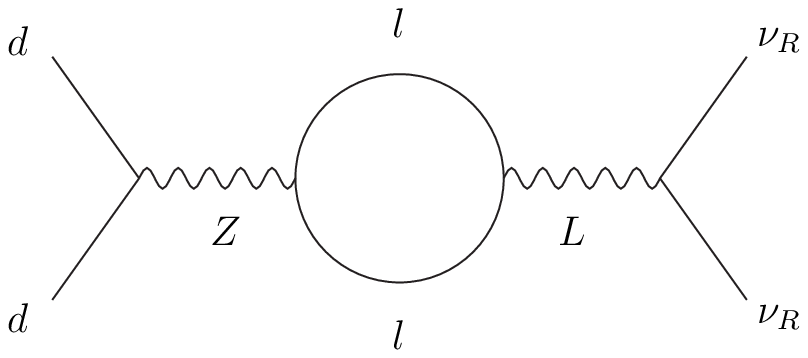}\end{center}
\end{figure}

This effective interaction can be mediated by the ``right'' gauge
boson (the one corresponding to the $U(1)$ gauge group in the right
brane) at tree level as well as through the one loop mixing between
the $Z$ and the ``leptonic'' gauge bosons as represented in
Fig.~\ref{supernova}. 
The details of course depend on the particular values of the masses
and couplings of the corresponding gauge bosons to the fermions. These
were discussed in~\cite{Ghilencea:2002da}. A good estimation is however
possible to make, taking into account the different possibilities. We
obtain the following (again conservative) bound from the tree level
process
\begin{equation}
M_s \gtrsim 5-10 \, \mathrm{TeV},
\end{equation}
whereas the loop-suppressed process gives
\begin{equation}
M_s \gtrsim 1 \, \mathrm{TeV}.
\end{equation}

Other processes like LEP constraints on contact interactions or the
effects of the extra $U(1)$'s on the $\rho$ parameter, lead to less
stringent bounds on the string scale of the order of the TeV.

\section{Conclusions}

Models with intersecting branes have many phenomenologically appealing
features. In order to further test their potential for fully realistic
models, we have considered their flavor structure. In this way we have
been able to compute in an unambiguous way their contributions to a
number of flavor violating and preserving operators. The resulting
bounds are very stringent
\begin{equation}
M_s \gtrsim 10^4\mbox{ TeV}.
\end{equation}
This feature will be generic in
any string model in which the different generations of fermions live
at separate points in the compact space. In that case, unless flavor
violations are strongly suppressed (an example of this situation
occurs in the warped case) for any reason, the high string scale
implied by them is difficult to make compatible with a stabilized
electroweak scale in non-supersymmetric models.

We should not be disappointed by these results. 
They should instead encourage us in the search of realistic
supersymmetric models for which much higher string and
compactification scales are stable against radiative corrections.

\section*{Acknowledgements}

It is a great pleasure to thank S.A. Abel, O. Lebedev and M. Masip for
collaboration and I. Navarro for useful discussions. This work has
been funded by PPARC.


\section*{References}


\begin{thebibliography}{99}

\bibitem{Berkooz:1996km}
M.~Berkooz, M.~R.~Douglas and R.~G.~Leigh,
Nucl.\ Phys.\ B {\bf 480} (1996) 265
[arXiv:hep-th/9606139].

\bibitem{Bachas:1995ik}
C.~Bachas,
arXiv:hep-th/9503030.



\bibitem{Ibanez:2001nd}
L.~E.~Iba\~nez, F.~Marchesano and R.~Rabad\'an,
JHEP {\bf 0111} (2001) 002
[arXiv:hep-th/0105155].


\bibitem{Blumenhagen:2000wh}
R.~Blumenhagen, L.~Goerlich, B.~Kors and D.~Lust,
JHEP {\bf 0010} (2000) 006
[arXiv:hep-th/0007024];
S.~Forste, G.~Honecker and R.~Schreyer,
Nucl.\ Phys.\ B {\bf 593} (2001) 127
[arXiv:hep-th/0008250];
G.~Aldazabal, S.~Franco, L.~E.~Iba\~nez, R.~Rabad\'an and A.~M.~Uranga,
JHEP {\bf 0102} (2001) 047
[arXiv:hep-ph/0011132];
G.~Aldazabal, S.~Franco, L.~E.~Iba\~nez, R.~Rabad\'an and A.~M.~Uranga,
J.\ Math.\ Phys.\  {\bf 42} (2001) 3103
[arXiv:hep-th/0011073].

\bibitem{Abel:stringreview}
S.~Abel and J.~Santiago,
J.\ Phys.\ G {\bf 30} (2004) R83
[arXiv:hep-ph/0404237].


\bibitem{Cvetic:2001tj}
M.~Cvetic, G.~Shiu and A.~M.~Uranga,
Phys.\ Rev.\ Lett.\  {\bf 87} (2001) 201801
[arXiv:hep-th/0107143];
M.~Cvetic, G.~Shiu and A.~M.~Uranga,
Nucl.\ Phys.\ B {\bf 615} (2001) 3
[arXiv:hep-th/0107166];
D.~Cremades, L.~E.~Iba\~nez and F.~Marchesano,
JHEP {\bf 0207} (2002) 009
[arXiv:hep-th/0201205];
R.~Blumenhagen, L.~Gorlich and T.~Ott,
JHEP {\bf 0301} (2003) 021
[arXiv:hep-th/0211059];
M.~Cvetic, I.~Papadimitriou and G.~Shiu,
arXiv:hep-th/0212177;
M.~Cvetic and I.~Papadimitriou,
Phys.\ Rev.\ D {\bf 67} (2003) 126006
[arXiv:hep-th/0303197];
M.~Cvetic, P.~Langacker and J.~Wang,
Phys.\ Rev.\ D {\bf 68} (2003) 046002
[arXiv:hep-th/0303208];
K.~Behrndt and M.~Cvetic,
Nucl.\ Phys.\ B {\bf 676} (2004) 149
[arXiv:hep-th/0308045].

\bibitem{Honecker:2003vq}
G.~Honecker,
Nucl.\ Phys.\ B {\bf 666} (2003) 175
[arXiv:hep-th/0303015].

\bibitem{Cremades:2003qj}
D.~Cremades, L.~E.~Iba\~nez and F.~Marchesano,
JHEP {\bf 0307} (2003) 038
[arXiv:hep-th/0302105].

\bibitem{Abel:2003fk}
S.~A.~Abel, M.~Masip and J.~Santiago,
JHEP {\bf 0304} (2003) 057
[arXiv:hep-ph/0303087].


\bibitem{Cvetic:2003ch}
M.~Cvetic and I.~Papadimitriou,
Phys.\ Rev.\ D {\bf 68} (2003) 046001
[arXiv:hep-th/0303083].

\bibitem{Abel:2003vv}
S.~A.~Abel and A.~W.~Owen,
Nucl.\ Phys.\ B {\bf 663} (2003) 197
[arXiv:hep-th/0303124]; 
S.~A.~Abel and A.~W.~Owen,
arXiv:hep-th/0310257.


\bibitem{Delgado:1999sv}
A.~Delgado, A.~Pomarol and M.~Quir\'os,
JHEP {\bf 0001} (2000) 030
[arXiv:hep-ph/9911252];
C.~D.~Carone,
Phys.\ Rev.\ D {\bf 61} (2000) 015008
[arXiv:hep-ph/9907362].



\bibitem{Abel:Lebedev:Santiago}
S.~A.~Abel, O.~Lebedev and J.~Santiago,
arXiv:hep-ph/0312157.




\bibitem{Langacker:2000ju}
P.~Langacker and M.~Plumacher,
Phys.\ Rev.\ D {\bf 62} (2000) 013006
[arXiv:hep-ph/0001204].


\bibitem{Demir:2003js}
D.~Demir, O.~Lebedev, K.~A.~Olive, M.~Pospelov and A.~Ritz,
arXiv:hep-ph/0311314;
O.~Lebedev and M.~Pospelov,
Phys.\ Rev.\ Lett.\  {\bf 89} (2002) 101801
[arXiv:hep-ph/0204359].


\bibitem{Ghilencea:2002da}
D.~M.~Ghilencea, L.~E.~Ibanez, N.~Irges and F.~Quevedo,
JHEP {\bf 0208}, 016 (2002);
D.~M.~Ghilencea,
Nucl.\ Phys.\ B {\bf 648}, 215 (2003).


\end{thebibliography}
\end{document}